**Sb concentration dependent structural and resistive properties of polycrystalline Bi-Sb alloys**

K. Malik[1], Diptasikha Das[1], D. Mondal[2], D. Chattopadhyay[2], A. K. Deb[3], S. Bandyopadhyay[1] and Aritra Banerjee[1,a)]

[1] Department of Physics, University of Calcutta, 92 A P C Road, Kolkata-700 009, India

[2] Dept. of Polymer Science and Technology, University of Calcutta, 92 A P C Road, Kolkata-700 009, India

[3] Department of Physics, Raiganj College (University College), Uttar Dinajpur, 733 134, West Bengal, India

**ABSTRACT**

Polycrystalline $Bi_{1-x}Sb_x$ alloys have been synthesized over a wide range of antimony concentration ($0.08 \leq x \leq 0.20$) by solid state reaction method. In depth structural analysis using X-Ray diffraction (XRD) and temperature dependent resistivity ($\rho$) measurement of synthesized samples have been performed. XRD data confirmed single phase nature of polycrystalline samples and revealed that complete solid solution is formed between bismuth and antimony. Rietveld refinement technique, utilizing MAUD software, has been used to perform detail structural analysis of the samples and lattice parameters of polycrystalline $Bi_{1-x}Sb_x$ alloys have been estimated. Lattice parameter and unit cell volume decreases monotonically with increasing antimony content. The variation of lattice parameters with antimony concentration depicts a distinct slope change at $x = 0.12$. Band gap ($E_g$) has been estimated from the thermal variation of resistivity data, with the 12% Sb content sample showing maximum $E_g$. It has been observed that, with increasing antimony concentration the transition from direct to indirect gap semiconductor is intimately related to the variation of the estimated lattice parameters. Band diagram for the polycrystalline $Bi_{1-x}Sb_x$ alloy system has also been proposed.

---

[a)] Author to whom correspondence should be addressed. Electronic mail: arbphy@caluniv.ac.in

## I. Introduction:

Thermoelectric materials have attracted much interest in recent years due to their applicability in power generation and electronic refrigeration.[1] Among the many developed thermoelectric materials, narrow gap semiconductors like bismuth-antimony (Bi-Sb) alloys have been presented as a highly attractive and remarkable thermoelectric material.[2,3] Apart from that, among semimetal and narrow gap semiconductors, bismuth-antimony (Bi-Sb) alloy have received special attention due to their interesting physical properties.[4-7] The band structure of Bi and Bi-Sb alloy is peculiar in nature and drastically depends on different physical parameters such as antimony concentration,[4,8,9] temperature,[10,11] pressure[5,12,13] and magnetic field.[14,15] Such peculiarities led to the observation of several exciting properties, viz., the existence of gapless state.[15,16] This further renewed interest in this material with the recent discoveries of topological insulator and dirac fermion.[17-19].

The group V elements, Bi and Sb are semimetals with $A_7$ type rhombohedral structure.[20,21] The semimetallic character of bismuth arises due to the small overlap of the valence band maximum at the T point with the conduction band minimum by 0.0184 eV at the L point of the brilloiun zone.[8,11] $Bi_{1-x}Sb_x$ alloy form a solid solution over the entire composition range.[6,21] The addition of Sb to Bi causes the $L_s$ and T bands to move down with respect to the $L_a$ band. At x = 0.04 the L bands invert [9,15,20] and at x ~ 0.07 the overlap between the hole T and $L_a$ bands disappears, resulting in semiconducting Bi-Sb alloy.[9,20] $Bi_{1-x}Sb_x$ alloy is semiconducting in the composition range x = 0.07-0.22 and depending on Sb content it becomes direct or indirect semiconductor. With the properties such as small band gap ($E_g$), high mobility and small effective masses, semiconducting Bi rich $Bi_{1-x}Sb_x$ alloy is potentially attractive for use as n type thermoelectric material.[3,20]

There have been several earlier reports on the structural and transport, viz, electrical, thermoelectric properties of $Bi_{1-x}Sb_x$ alloy. Lin et al reported the thermoelectric properties of $Bi_{1-x}Sb_x$ alloy nanowires.[22] But most of the reports are on single crystalline[18,20] and thin films[9,23,24] of Bi-Sb alloy. Based on the resistivity (ρ) and thermoelectric power data, different research groups have estimated the band diagram of the single crystalline and thin film $Bi_{1-x}Sb_x$ samples. But there are some discrepancies in the reported band diagrams. However, single crystals are weak in mechanical strength and require dedicated techniques for synthesis. Textured materials also provide enough mechanical strength and might be beneficiary in samples with anisotropic properties.[25] It is note worthy to mention that, Bi-Sb alloys exhibit highly anisotropic properties [8,15,20] and thus textured samples might give rise to larger thermoelectric power factors and figures of merits, as observed by D. Kenfaui for oxide thermoelectric material[25]. However synthesis of textured materials also requires specialized techniques[25], as compared to bulk polycrystalline samples. Hence, the use of polycrystalline materials might be more suitable for practical applications. But there are very limited reports on transport property study of polycrystalline semiconducting Bi-Sb alloys.[3,5,6,13] H. Kitagawa et al reported the thermoelectric properties of polycrystalline $Bi_{1-x}Sb_x$ alloy system.[6] Very recently S. Dutta et al reported the effect of pressure and temperature on ρ and thermopower of polycrystalline Bi-Sb samples.[5,13] Some efforts also have been given to reveal the structural properties of rhombohedral $Bi_{1-x}Sb_x$ alloys.[7,8,23,26] But none of them have used Rietveld refinement techniques. It should be mentioned that in these group of samples, Souza et al only recently employed Rietveld refinement method for structural analysis of another important thermoelectric material having rhombohedral structure, $Bi_2Te_3$.[27]

In the present paper, we report the detailed structural analysis and electrical resistivity (ρ) study of the polycrystalline $Bi_{1-x}Sb_x$ (0.08 ≤ x ≤ 0.20) alloys synthesized by solid state reaction method. From the thermal variation of ρ data, $E_g$ for these narrow gap semiconductors have been estimated. Further, the ρ data have been used to construct the band diagram, i.e., variation of energy bands near the Fermi level of this polycrystalline $Bi_{1-x}Sb_x$ alloy system. Structural analysis has been performed by Rietveld refinement method using the X-Ray Diffraction (XRD) data of the synthesized Bi-Sb alloys. It has been mentioned earlier that, based on antimony (x) content $Bi_{1-x}Sb_x$ alloy can be direct or indirect semiconductor. In this article, we have demonstrated that, the transition from direct to indirect gap semiconductor is accompanied by a change in estimated structural parameters, e.g., lattice parameters.

**II. Experimental:**

The $Bi_{1-x}Sb_x$ ($0.08 \leq x \leq 0.20$) alloys were synthesized by conventional solid-state reaction method.[28] Stoichiometric amount of Bi and Sb powders (each of purity 99.999%; Alfa Aesar, UK) have been weighed, mixed and loaded into a quartz ampoule, sealed under vacuum at $10^{-3}$ Pa. The quartz ampoules containing the powder mixtures are then sintered at $700^0$C for 10 h. The structural characterization of the synthesized $Bi_{1-x}Sb_x$ samples has been carried out using powder x-ray diffractometer [Model: X'Pert PRO (PANalytical)] with Cu-$K_\alpha$ radiation. All the XRD measurements are performed in the range of $20^0 \leq 2\theta \leq 80^0$ in θ-2θ geometry. Rietveld refinement technique was used to perform in depth structural analysis of the synthesized samples. Utilizing the MAUD software,[29] the Rietveld method was used to refine the structural parameters from the XRD patterns of the obtained polycrystalline $Bi_{1-x}Sb_x$ alloy. Standard Si sample was used for measuring the instrumental profile.[30] The electrical resistivity as a function of temperature (ρ−T) of all the samples was measured using conventional four-probe technique.

### III. Results and Discussion:

XRD patterns of the synthesized $Bi_{1-x}Sb_x$ ($0.08 \leq x \leq 0.20$) samples are shown in Figure 1. The XRD patterns reveal that, all the samples are single phase in nature. The variation of the most intense (012) diffraction peak with 2θ, for different Sb concentration is demonstrated in Figure 1(inset). All the (012) peaks of the $Bi_{1-x}Sb_x$ alloy shift to high angle side with increasing antimony concentration. The atomic radii of antimony (1.45Å) are less than that of bismuth (1.60Å) and Bi rich Bi-Sb alloys follow Vegard's law,[6] thus it is quite expected that, with increasing Sb concentration, the diffraction peaks should shift to higher angle.[6] This further confirms that, a complete solid solution has been formed between Bi and Sb and synthesized $Bi_{1-x}Sb_x$ alloys are single phase in nature. We have also determined the full width at half maxima (FWHM) of all the synthesized $Bi_{1-x}Sb_x$ samples for the most intense peak. The results are shown in Figure 2a. With increasing Sb, FWHM is observed to increase gradually, implying poorer crystalline quality of the Bi-Sb alloy with increasing antimony concentration. This is a typical alloying behaviour, which arises due to the difference in the lattice constant as well as the atomic radii of Bi and Sb.[23]

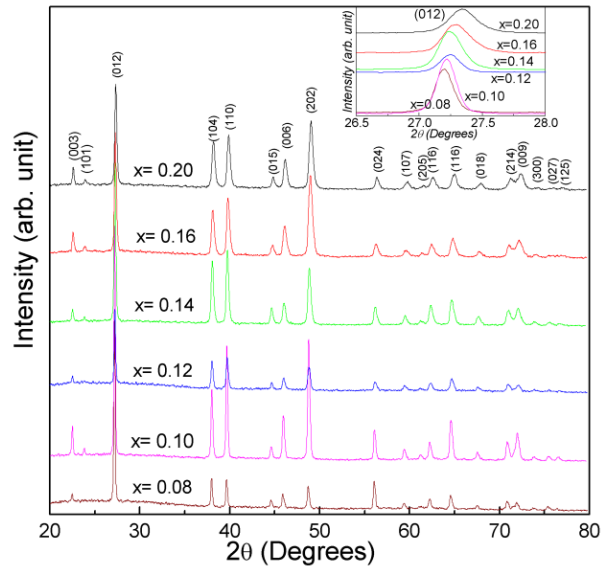

**FIG. 1.** XRD pattern of $Bi_{1-x}Sb_x$ ($0.08 \leq x \leq 0.20$) alloys. Inset shows (012) peak of rhombohedral $Bi_{1-x}Sb_x$ alloys.

Bi-Sb alloy crystallizes in rhombohedral $A_7$ type structure.[20,21] In order to analyse the effect of Sb concentration on the structure of the $Bi_{1-x}Sb_x$ alloys, we have performed in-depth structural analysis, using Rietveld refinement technique of our samples. This further helps us in checking the quality of our synthesized samples as well as confirms the single phase nature of all the synthesized $Bi_{1-x}Sb_x$ alloys. It is noteworthy to mention that there are some reports on the structural analysis of $Bi_{1-x}Sb_x$ single crystals, thin films[8,23] and polycrystalline Bi-Sb alloys.[7,26,31] But, till date no effort was given to reveal the effect of Sb concentration on the structural property of polycrystalline $Bi_{1-x}Sb_x$ alloys using Rietveld refinement method. Very recently, only Souza et al reported the structural property of polycrystalline $Bi_2Te_3$, another well known thermoelectric material of rhombohedral structure, using Rietveld method.[27] The experimental XRD pattern and theoretical XRD curve, as obtained after Rietveld refinement, for all the synthesized Bi-Sb alloys are presented in Figure S1 [See the supplementary information for Figure S1 (*page 18-19*)]. The goodness of fit (GoF) or $\chi^2$ values obtained after refinements, given in Table 1, signifies that high quality structural analysis by Rietveld refinement technique has been performed. It is noteworthy to mention that, we have performed the refinement using the atomic positions and substitutions of the synthesized Bi-Sb alloy and the corresponding parameters like site occupancy, atomic positions, reliability parameters ($R_w$, $R_b$, $R_{exp}$) etc., obtained after refinement has been included in supplementary information (Figure S1). Space group $R\bar{3}m$ and point group $D_{3d}$ was used for refinement. The description of rhombohedral Bi-Sb alloys is commonly expressed in terms of equivalent hexagonal unit cell.[8,26,32] It should be mentioned here that, after proper coordinate transformation, the rhombohedral crystal structure can be described with a hexagonal unit cell.[32] For hexagonal unit cell $\alpha=\beta=90^0$, $\gamma=120^0$ and the corresponding lattice parameters, $a_H$ and $c_H$, obtained upon refinement has been found to decrease with increasing Sb concentration (Table-I). Similar results has also been reported by other groups for Bi-Sb thin-films or single crystal.[8,23] Our Rietveld refinement data also reveals that, the estimated hexagonal unit cell volume of the $Bi_{1-x}Sb_x$ alloys decreases with increasing antimony (x), as given in Table-I. The atomic radii of Sb (1.45 Å) are smaller than that of Bi (1.60 Å).[33] With increasing Sb concentration (x), more and more bismuth atoms are replaced by antimony atoms in unit cells of Bi-Sb alloys. Thus, such decrease of lattice parameters (both $a_H$ and $c_H$) and hence the corresponding shrinkage of unit cell volume of the $Bi_{1-x}Sb_x$ alloys with increasing Sb concentration is expected. Figure 2b shows a plot of lattice parameters (both $a_H$ and $c_H$) versus Sb concentration for our synthesized $Bi_{1-x}Sb_x$ samples.

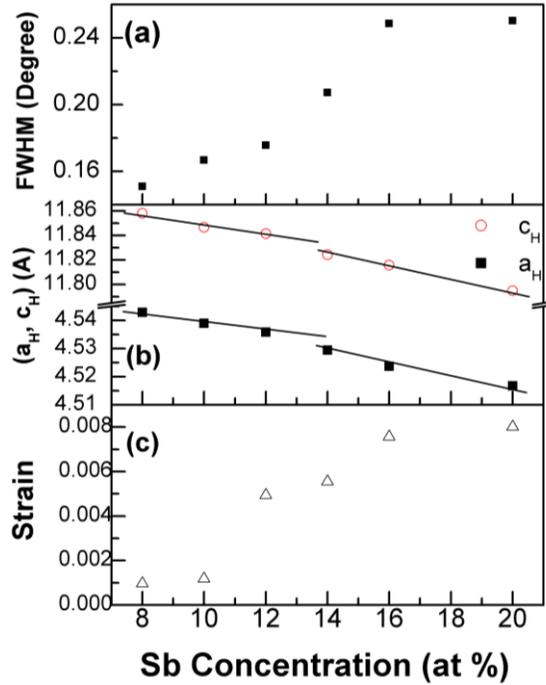

FIG. 2. Antimony concentration dependence of (a) FWHM of (012) peak (b) hexagonal lattice parameters ($a_H$, $c_H$) and (c) Strain (using Williamson-Hall method) as obtained on synthesized $Bi_{1-x}Sb_x$ samples.

Careful investigation of Figure 2b clearly reveals that, the variation of lattice parameters with antimony concentration shows a slope change for x > 0.12. For lower concentration of antimony (x), i.e., for x=0.08 to x=0.12, both $a_H$ and $c_H$ slowly decreases with Sb concentration. But for higher Sb concentration, i.e., for x $\geq$ 0.14, $a_H$ and $c_H$ decreases rather sharply. It should be mentioned here that, antimony concentration dependent band gap ($E_g$) variation, as obtained from ρ–T data [Figure 3], also shows a slope change at x = 0.12 [Figure 4, discussed below].

Further from our XRD patterns, we have also estimated the strain developed in the synthesized polycrystalline Bi-Sb samples and its grain size, employing Williamson-Hall method [34] as well as by refining the size-strains using harmonic deconvolutions in the Rietveld code. The estimated values of strain and grain size, as obtained from both the method, follows similar trend (Table-II). The Williamson-Hall plots for all the synthesized $Bi_{1-x}Sb_x$ samples have been included in Figure S2 [See the supplementary information for Figure S2 (*page 20*)]. The estimated strain values, plotted in Figure 2c, increases with increasing antimony concentration. The difference in the atomic sizes of Bi and Sb leads to the shrinkage of unit cell volume in $Bi_{1-x}Sb_x$ alloy system. Thus the size mismatch of Bi and Sb atoms in turn develops some strain in $Bi_{1-x}Sb_x$ unit cell, which increases with increasing antimony concentration.

The ρ−T curve of the polycrystalline $Bi_{1-x}Sb_x$ samples for various Sb compositions are shown in Figure 3. As the Sb concentration increases, ρ increases for the synthesized $Bi_{1-x}Sb_x$ alloys. This is related to the increased impurity scattering in the system with increased Sb doping. In addition, the grain size also decreases (Table-II), which in turn increases grain boundary scattering, with increasing Sb concentration. Thus the increased grain boundary scattering also plays a significant role in the observed increase of ρ with

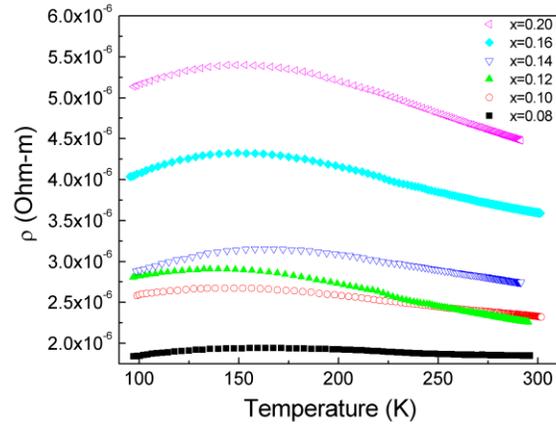

**FIG. 3.** Temperature dependence electrical resistivity (ρ) of $Bi_{1-x}Sb_x$ alloys.

increasing Sb content in $Bi_{1-x}Sb_x$ system. Further, in Figure 3, the measured ρ of the $Bi_{1-x}Sb_x$ alloys exhibits non-monotonic temperature dependence, with a maximum at around 150 K. Temperature at which ρ value shows a maximum has been designated as peak temperature, $T_p$. The exact values of $T_p$ depend on the antimony content and its variation with Sb concentration (x) has been plotted in Figure 4. Figure 4 also includes the variation of $E_g$ with Sb content (discussed below) and it has been found that Sb concentration dependent change of $T_p$ and $E_g$ is very much correlated. The ρ−T curve [Figure 3] depicts that, ρ first begins to increase with increasing temperature upto $T_p$, then decreases with further increase of temperature. S. Cho et al and Y. Lin et al had also reported non-monotonic temperature dependence of ρ for $Bi_{1-x}Sb_x$ thin films and nano wires, respectively.[9,22] Similar semi-metal to semiconductor transition has also been reported earlier for single crystalline Bi-Sb alloy system.[4,8,20] However, Lenoir et al reported the transition at much lower temperature, around 50K.[4,20] It should be pointed out that our measurement is on polycrystalline samples, where grain boundary and other related scattering events also plays a significant role. This might lead to higher $T_p$ obtained in our measurement as compared to those reported by Lenoir et al.[4,20] This low temperature metallic behaviour of electrical resistivity has been explained by Lenoir et al in terms of so called hydrogeniod model.[4,20] With increasing temperature, due to increased thermal energy, the intrinsic density of carriers in the system increases. For temperature above $T_p$, this intrinsic carrier density subsequently becomes greater than the density of impurities. This leads to the observed temperature dependence of semiconductors, showing a decrease in ρ with increasing temperature above $T_p$. Bi-Sb alloy system is a narrow band gap semiconductor. The electrical resistivity in the semiconducting range (above $T_p$) follows exponential behaviour and the corresponding band gap ($E_g$) was estimated from the ρ−T data using the relation (for $T \geq T_p$):

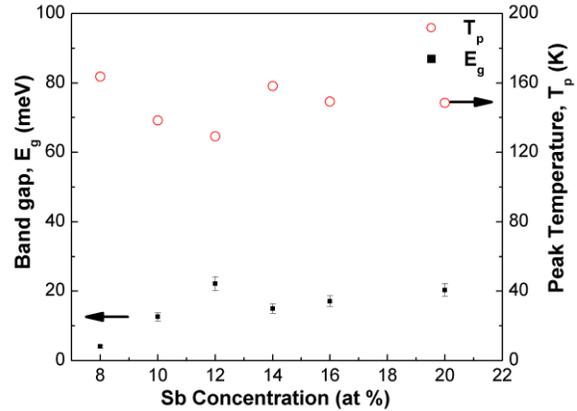

FIG. 4. Evolution of peak temperature ($T_p$) and band gap ($E_g$) as a function of Sb concentration for Bi rich Bi-Sb alloys.

$$\rho = \rho_0 \exp\left(\frac{E_g}{2k_B T}\right) \qquad (1)$$

where, $\rho_0$ is a constant. The estimated values of $E_g$, plotted in figure 4, have been used to describe qualitatively the evolution of band structure as a function of antimony composition. Similar attempts have been made earlier for single crystalline samples[4,8,20] as well as for epitaxial thin films,[9] but there are not of much report for polycrystalline samples.[5,15] The semiconducting behaviour in bulk system occurs at alloy composition between 7% and 22% Sb and Lenoir et al obtained a maximum $E_g$ (21 meV) at x = 0.15.[4,20] Whereas, A. L Jain reported the highest $E_g$ for x=0.12.[8] The Sb concentration for the maximum gap shifts to lower Sb concentration of 9% for $Bi_{1-x}Sb_x$ thin films grown on CdTe (111), as reported by S. Cho et al.[9] On the other hand in the present work for the synthesized polycrystalline samples, the maximum $E_g$ has been observed for 12% Sb concentration, in accordance to those reported by A L Jain.[8]

According to the energy band model proposed by Blount and Cohen[35] the band structure of bismuth, consists of a pair of light mass bands ($L_a$ and $L_s$) at six symmetrically related positions in k space, the upper of which is occupied by electrons and a heavy mass hole band (T) at two positions in k space.[5,8,35] In antimony, the pockets of electrons are also located at the L points of the Brillouin zone, where as holes are located at the H points.[20] The effect of Sb addition causes the $L_s$ and T bands to move downward with respect to the $L_a$ band. At x = 0.07 the overlap between the hole T and $L_a$ bands disappears, resulting in an indirect gap semiconductor for $x \cong 0.07$. With further increase of Sb concentration, the heavy mass band (T) further shifts downward, with respect to the light mass band and the $Bi_{1-x}Sb_x$ alloy system transforms to a direct gap ($L_a$-$L_s$) narrow band semiconductor. Thus the polycrystalline $Bi_{1-x}Sb_x$ alloy samples reported here, is a direct gap semiconductor for lower Sb concentration i.e., for $0.08 \leq x \leq 0.12$. Figure 4 demonstrates that for increasing Sb concentration between 8 % and 12 %, $E_g$ increases. This signifies that, with increasing antimony content in the range 0.08 < x <

0.12, T band shifts downward further, which in turn increases the estimated direct gap ($L_a$-$L_s$). For increasing Sb content beyond 12 %, the hole H band lies above $L_s$ and we have an indirect semiconductor ($L_a$-H). From Figure 4 it has been further observed that, increasing antimony concentration for x > 0.12, decreases this estimated indirect $E_g$. Actually for $0.14 \leq x \leq 0.20$, the evolution of H band is such that, the effective $E_g$ ($L_a$-H) is indirect and decreases with increasing antimony concentration. As discussed above, Figure 2b clearly indicated that, the variation of lattice parameter with antimony concentration also shows a distinct slope change at x = 0.12. Thus the concurrent variation of $E_g$ with Sb concentration and the concentration dependent variation of $a_H$ and $c_H$, clearly demonstrate that with increasing antimony concentration the transition from direct to indirect gap semiconductor in $Bi_{1-x}Sb_x$ alloy system is intimately related to the variation of the estimated lattice parameters. Based on the structural and the ρ−T presented here, we propose a band diagram, i.e., variation of energy bands near the Fermi level for the polycrystalline $Bi_{1-x}Sb_x$ alloy system. The proposed band diagram is given in Figure 5. The corresponding E-k diagram (schematic) for different $Bi_{1-x}Sb_x$ ($0.08 \leq x \leq 0.20$) samples discussed here has been presented in Figure S3 [See the supplementary information for Figure S3 (*page 21*)]. It should be noted that, the band diagram proposed here is mainly based on equation 1, which suggests that the bands are parabolic with the same density of state. But there are multiple hole bands (L, T and H), among them $L_s$ is 'light' hole band where as T and H bands are 'heavy' hole bands, i.e., bands with greater density of states. The band which makes dominant contribution to thermal variation of electrical resistivity is a complex interplay between position of respective bands, density of states, mobility and temperature. Also, equation 1 assumes that, carriers are scattered by acoustical phonons. Hence, it is tough to apply the above formula directly for polycrystalline samples, where other scattering mechanisms like grain boundary scattering also plays a significant role. Despite these approximations, for a qualitative understanding of Bi-Sb alloy system the use of this simple model is quite justified for describing the evolution of the band structure as a function of composition.[9,20]

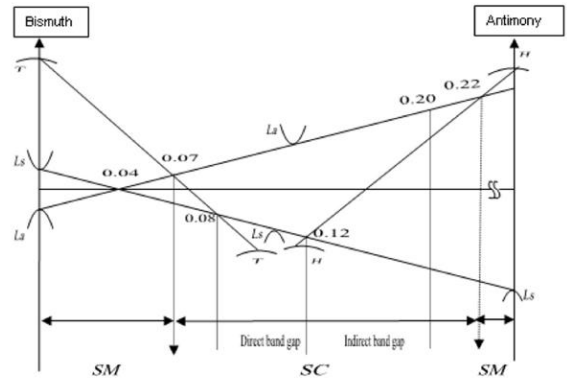

**FIG. 5.** Schematic representation of energy bands near the Fermi level for $Bi_{1-x}Sb_x$ alloys as a function of Sb composition (x). Depending on Sb concentration, Bi-Sb alloys can be semi-metal (SM) or Semiconductor (SC), which is indicated in the diagram. For simplicity, the L, T and H point bands are drawn one on top of the others.

## IV. Conclusion:

Structural analysis and measurement of thermal variation of resistivity data have been performed on polycrystalline $Bi_{1-x}Sb_x$ ($0.08 \leq x \leq 0.20$) alloys, synthesized by solid state reaction method. XRD data revealed that the samples are single phase in nature. Rietveld refinement technique has been employed for in-depth structural analysis of rhombohedral $Bi_{1-x}Sb_x$ samples. The description of a rhombohedral structure is commonly expressed in terms of equivalent hexagonal unit cell. Rietveld refinement data shows that, hexagonal lattice parameter $a_H$ and $c_H$, decreases systematically with increasing antimony concentration. The Sb concentration dependent variation of $a_H$ and $c_H$ demonstrates a distinct slope change at x = 0.12. The electrical resistivity increases for the synthesized $Bi_{1-x}Sb_x$ alloys with increase of Sb concentration (x). In addition, the measured ρ-T curve of the synthesized samples exhibits non-monotonic temperature dependence with a maximum ρ around 150 K. High temperature semiconducting data has been analysed and $E_g$ for these series of narrow gap semiconductors has been estimated. The alloy with 12 % Sb concentration shows highest $E_g$ of 21 meV. It has also been demonstrated that, the $Bi_{1-x}Sb_x$ alloys with $0.08 \leq x \leq 0.12$ is direct gap semiconductor, where as for $0.14 \leq x \leq 0.20$ it becomes indirect gap semiconductor. It can be concluded that the transition from direct to indirect gap semiconductor is depicted by a distinct slope change in the plotted lattice parameter (both $a_H$ and $c_H$) vs Sb concentration curve. Further, based on the ρ-T data, band diagram for the polycrystalline $Bi_{1-x}Sb_x$ alloy system has been proposed.

**Acknowledgement:**

This work is supported by Department of Science and Technology (DST), Govt. of India and UGC, Govt. of India in the form of sanctioning research project, reference no. SF/FTP/PS-25/2009 and 39-990/2010(SR), respectively. Author KM is thankful to University Grants Commission (UGC) for providing him Junior Research Fellowship and author DD is grateful to DST, Govt of India for providing financial assistance in form of Junior Research fellowship through project no. SF/FTP/PS-25/2009.

**Table-I**. Structural parameters of Hexagonal unit cell of $Bi_{1-x}Sb_x$ ($0.08 \leq x \leq 0.20$) alloy for different Sb concentrations (x), obtained by Rietveld Refinement technique using MAUD software along with the corresponding Goodness of fit (GoF) or $\chi^2$ value.

| x | Hexagonal lattice parameter | | | GoF or $\chi^2$ |
|---|---|---|---|---|
| | $a_H$ (Å) | $c_H$ (Å) | Volume (Å$^3$) | |
| 0.08 | $4.5429 \pm 2.7 \times 10^{-4}$ | $11.8579 \pm 1.1 \times 10^{-3}$ | $211.9294 \pm 5.3 \times 10^{-3}$ | 1.138 |
| 0.10 | $4.5390 \pm 1.2 \times 10^{-4}$ | $11.8465 \pm 6.7 \times 10^{-4}$ | $211.3625 \pm 3.1 \times 10^{-3}$ | 1.334 |
| 0.12 | $4.5358 \pm 2.2 \times 10^{-4}$ | $11.8414 \pm 1.2 \times 10^{-3}$ | $210.9741 \pm 5.5 \times 10^{-3}$ | 1.066 |
| 0.14 | $4.5294 \pm 1.5 \times 10^{-4}$ | $11.8243 \pm 8.1 \times 10^{-4}$ | $210.0755 \pm 3.7 \times 10^{-3}$ | 1.199 |
| 0.16 | $4.5237 \pm 4.3 \times 10^{-4}$ | $11.8157 \pm 1.9 \times 10^{-3}$ | $209.3947 \pm 9.0 \times 10^{-3}$ | 1.389 |
| 0.20 | $4.5168 \pm 2.2 \times 10^{-4}$ | $11.7950 \pm 1.4 \times 10^{-3}$ | $208.3903 \pm 6.3 \times 10^{-3}$ | 1.292 |

**Table-II**. The strain and grain size values estimated from both Williamson-Hall (WH) method and Reitveld refinement technique for all synthesized $Bi_{1-x}Sb_x$ samples.

| Sb content (at %) | Strain estimated from | | Grain size estimated from | |
|---|---|---|---|---|
| | WH method | Rietveld Refinement | WH method (nm) | Rietveld Refinement (nm) |
| 08 | $0.98 \times 10^{-3} \pm 5.26 \times 10^{-5}$ | $0.68 \times 10^{-3} \pm 5.2 \times 10^{-5}$ | $455 \pm 50$ | $444 \pm 56$ |
| 10 | $1.05 \times 10^{-3} \pm 8.28 \times 10^{-5}$ | $1.26 \times 10^{-3} \pm 3.5 \times 10^{-5}$ | $402 \pm 32$ | $375 \pm 30$ |
| 12 | $4.94 \times 10^{-3} \pm 3.3 \times 10^{-4}$ | $1.59 \times 10^{-3} \pm 7.2 \times 10^{-5}$ | $305 \pm 60$ | $303 \pm 56$ |
| 14 | $5.55 \times 10^{-3} \pm 3.4 \times 10^{-4}$ | $2.25 \times 10^{-3} \pm 4.1 \times 10^{-5}$ | $167 \pm 22$ | $283 \pm 29$ |
| 16 | $7.57 \times 10^{-3} \pm 4.4 \times 10^{-4}$ | $3.17 \times 10^{-3} \pm 8.6 \times 10^{-5}$ | $148 \pm 12$ | $237 \pm 11$ |
| 20 | $8.01 \times 10^{-3} \pm 3.9 \times 10^{-4}$ | $3.16 \times 10^{-3} \pm 6.3 \times 10^{-5}$ | $108 \pm 16$ | $170 \pm 18$ |

**Supplementary Information**

**FIGURE S1:** Rietveld refinement pattern and the corresponding refinement parameters obtained using MAUD software for $Bi_{1-x}Sb_x$ alloys for: (a) x=0.08, (b) x=0.10, (c) x=0.12, (d) x=0.14, (e) x= 0.16, (f) x=0.20. The values given in bracket represent corresponding errors.

(a) 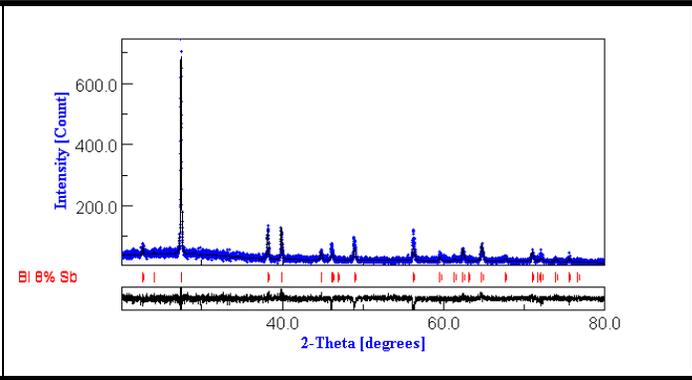

| Phase | $Bi_{0.92}Sb_{0.08}$ [ $R\bar{3}m$ ] |
|---|---|
| $Bi_{Occup.}$ | Bi1: 0.92 |
| $Sb_{Occup.}$ | Sb1: 0.08 |
| $Bi_x/Bi_y/Bi_z$ | 0.0 / 0.0/ 0.2328(3.0x10$^{-4}$) |
| $Sb_x/Sb_y/Sb_z$ | 0.0 / 0.0/ 0.2328(equal Bi) |
| $B_{iso\ Bi/Sb}$ | Bi1: 0.847 (0.133) Sb1: 0.847 (equal Bi) |
| $R_w$ (%) | 21.788 |
| $R_b$(%) | 16.675 |
| $R_{exp}$(%) | 19.147 |
| GoF or $\chi^2$ | 1.138 |

(b) 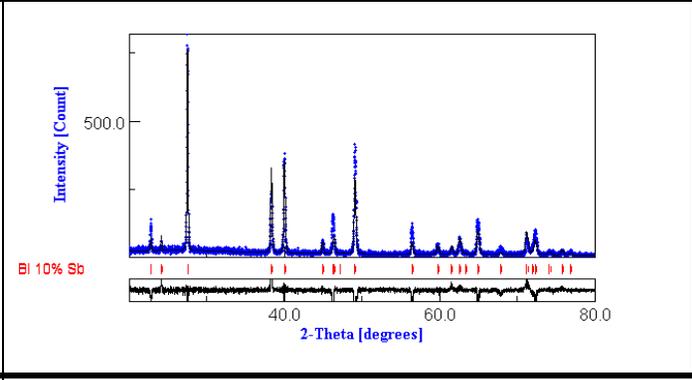

| Phase | $Bi_{0.90}Sb_{0.10}$ [ $R\bar{3}m$ ] |
|---|---|
| $Bi_{Occup.}$ | Bi1: 0.90 |
| $Sb_{Occup.}$ | Sb1: 0.10 |
| $Bi_x/Bi_y/Bi_z$ | 0.0 / 0.0/ 0.2303(1.8x10$^{-4}$) |
| $Sb_x/Sb_y/Sb_z$ | 0.0 / 0.0/ 0.2303(equal Bi) |
| $B_{iso\ Bi/Sb}$ | Bi1: 0.531 (0.092) Sb1: 0. 531 (equal Bi) |
| $R_w$ (%) | 25.519 |
| $R_b$(%) | 21.162 |
| $R_{exp}$(%) | 19.130 |
| GoF or $\chi^2$ | 1.334 |

(c) 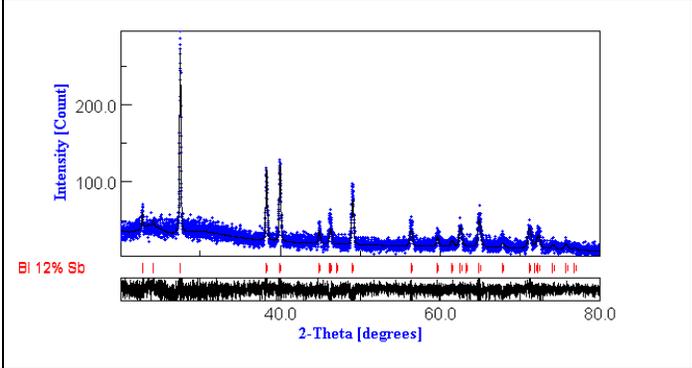

| Phase | $Bi_{0.88}Sb_{0.12}$ [ $R\bar{3}m$ ] |
|---|---|
| $Bi_{Occup.}$ | Bi1: 0.88 |
| $Sb_{Occup.}$ | Sb1: 0.12 |
| $Bi_x/Bi_y/Bi_z$ | 0.0 / 0.0/ 0.2314(3.3x10$^{-4}$) |
| $Sb_x/Sb_y/Sb_z$ | 0.0 / 0.0/ 0.2314(equal Bi) |
| $B_{iso\ Bi/Sb}$ | Bi1: 1.792 (0.150) Sb1: 1.792 (equal Bi) |
| $R_w$ (%) | 21.000 |
| $R_b$(%) | 16.351 |
| $R_{exp}$(%) | 19.701 |
| GoF or $\chi^2$ | 1.066 |

(d) 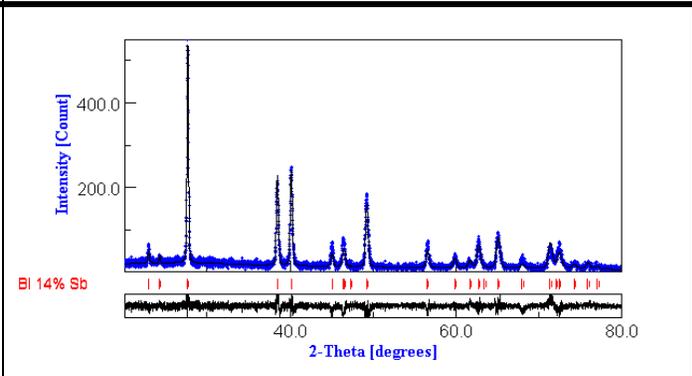

| Phase | $Bi_{0.86}Sb_{014}$ [ $R\bar{3}m$ ] |
|---|---|
| $Bi_{Occup.}$ | Bi1: 0.86 |
| $Sb_{Occup.}$ | Sb1: 0.14 |
| $Bi_x/Bi_y/Bi_z$ | 0.0 / 0.0/ 0.2323(1.8x10$^{-4}$) |
| $Sb_x/Sb_y/Sb_z$ | 0.0 / 0.0/ 0.2323(equal Bi) |
| $B_{iso\ Bi/Sb}$ | Bi1: 1.439 (0.087) Sb1: 1.439 (equal Bi) |
| $R_w$ (%) | 23.301 |
| $R_b$(%) | 18.541 |
| $R_{exp}$(%) | 19.434 |
| GoF or $\chi^2$ | 1.199 |

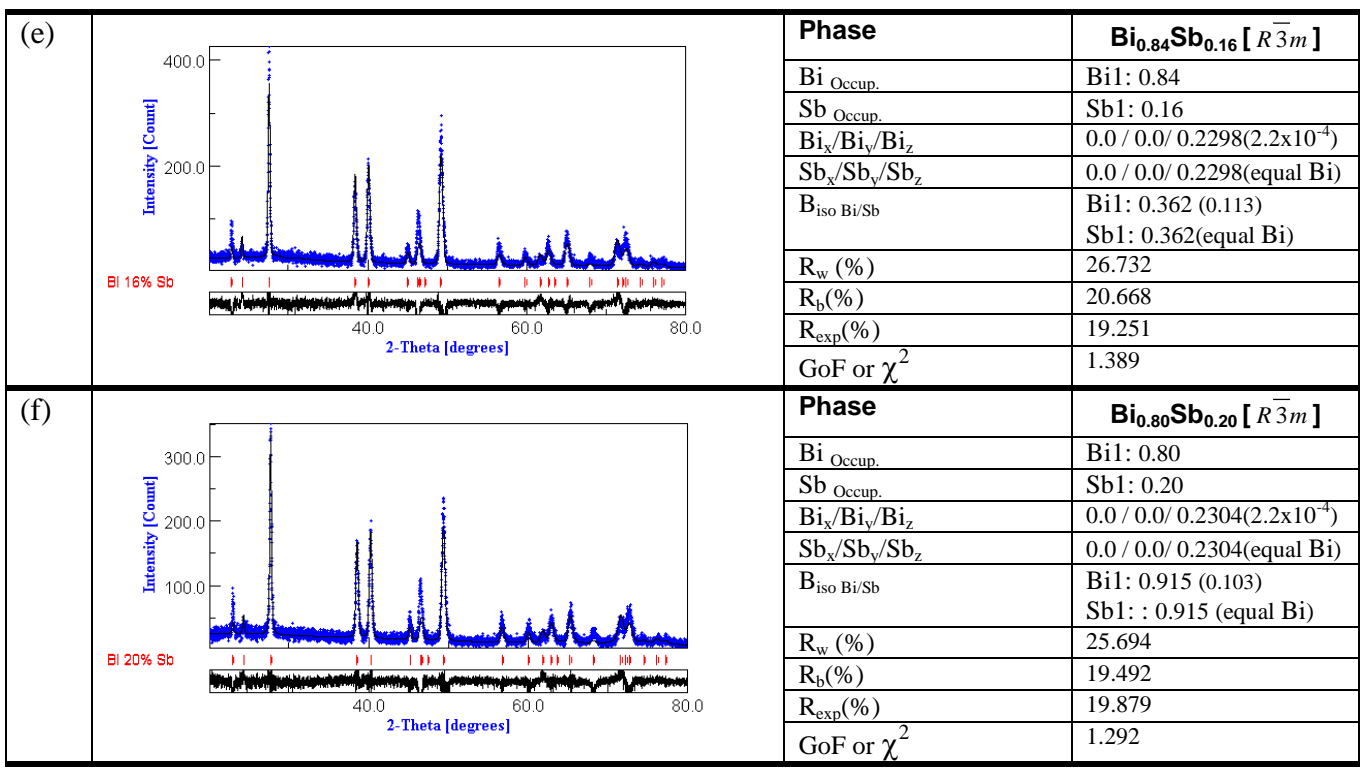

| | | Phase | Bi$_{0.84}$Sb$_{0.16}$ [ $R\bar{3}m$ ] |
|---|---|---|---|
| (e) | | Bi $_{Occup.}$ | Bi1: 0.84 |
| | | Sb $_{Occup.}$ | Sb1: 0.16 |
| | | Bi$_x$/Bi$_y$/Bi$_z$ | 0.0 / 0.0/ 0.2298(2.2x10$^{-4}$) |
| | | Sb$_x$/Sb$_y$/Sb$_z$ | 0.0 / 0.0/ 0.2298(equal Bi) |
| | | B$_{iso\ Bi/Sb}$ | Bi1: 0.362 (0.113)  Sb1: 0.362(equal Bi) |
| | | R$_w$ (%) | 26.732 |
| | | R$_b$(%) | 20.668 |
| | | R$_{exp}$(%) | 19.251 |
| | | GoF or $\chi^2$ | 1.389 |
| (f) | | Phase | Bi$_{0.80}$Sb$_{0.20}$ [ $R\bar{3}m$ ] |
| | | Bi $_{Occup.}$ | Bi1: 0.80 |
| | | Sb $_{Occup.}$ | Sb1: 0.20 |
| | | Bi$_x$/Bi$_y$/Bi$_z$ | 0.0 / 0.0/ 0.2304(2.2x10$^{-4}$) |
| | | Sb$_x$/Sb$_y$/Sb$_z$ | 0.0 / 0.0/ 0.2304(equal Bi) |
| | | B$_{iso\ Bi/Sb}$ | Bi1: 0.915 (0.103)  Sb1: : 0.915 (equal Bi) |
| | | R$_w$ (%) | 25.694 |
| | | R$_b$(%) | 19.492 |
| | | R$_{exp}$(%) | 19.879 |
| | | GoF or $\chi^2$ | 1.292 |

**FIGURE S2:** Williamson-Hall plots for different $Bi_{1-x}Sb_x$ ($0.08 \leq x \leq 0.20$) samples, where $\theta$ is the Bragg angle and $B_r$ is the Full Width at Half Maxima (FWHM) of the XRD peaks due to the combined effects of crystallite size and lattice strain. It should be mentioned that, if FWHM of the observed X-Ray peak is $B_0$ and FWHM due to the instrumental broadening is $B_i$, then $B_r = B_0 - B_i$

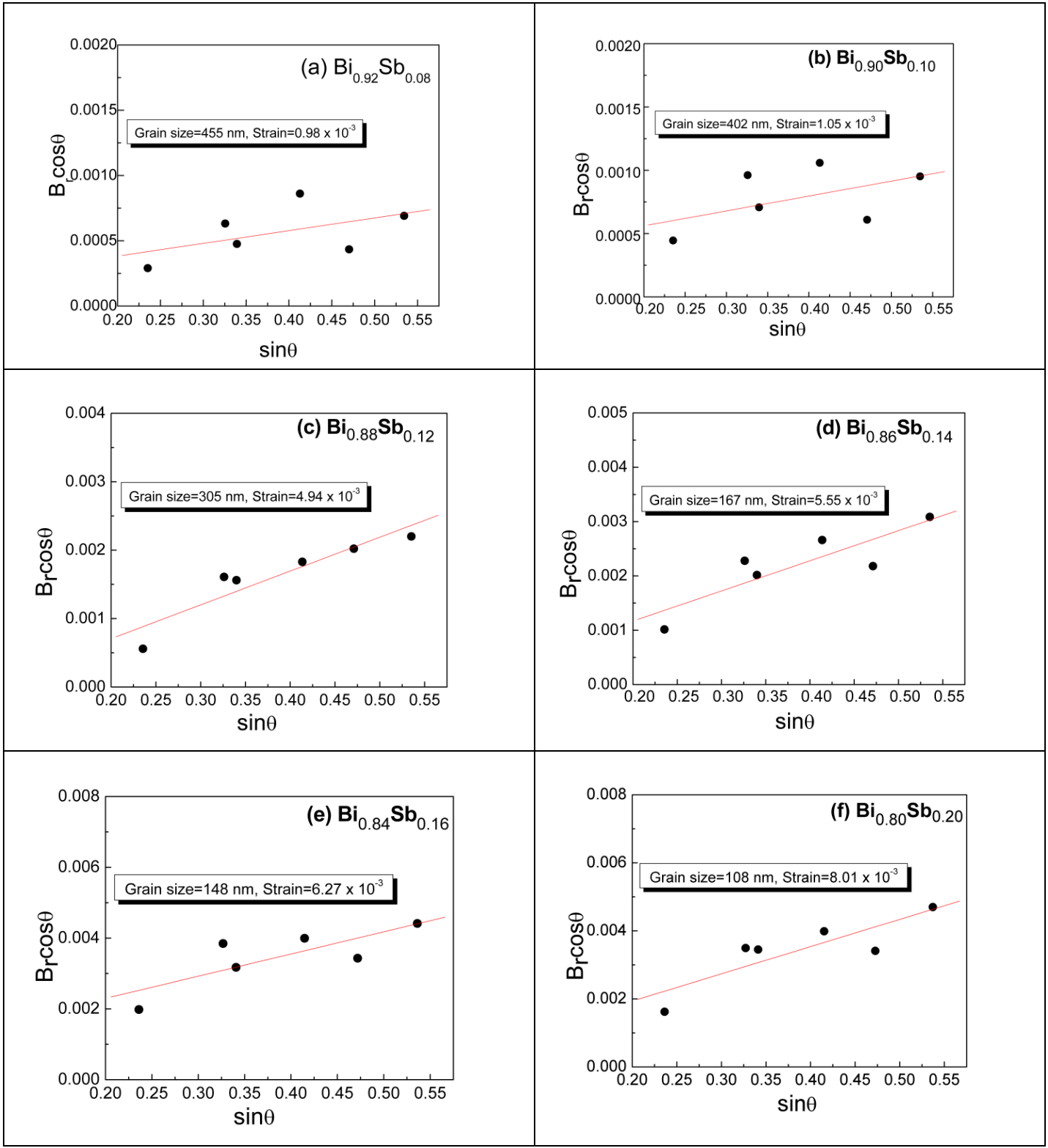

**FIGURE S3.** Schematic representation of $E$ vs. $k$ diagram for $Bi_{1-x}Sb_x$ ($0.08 \leq x \leq 0.20$) alloys for different Sb concentration.

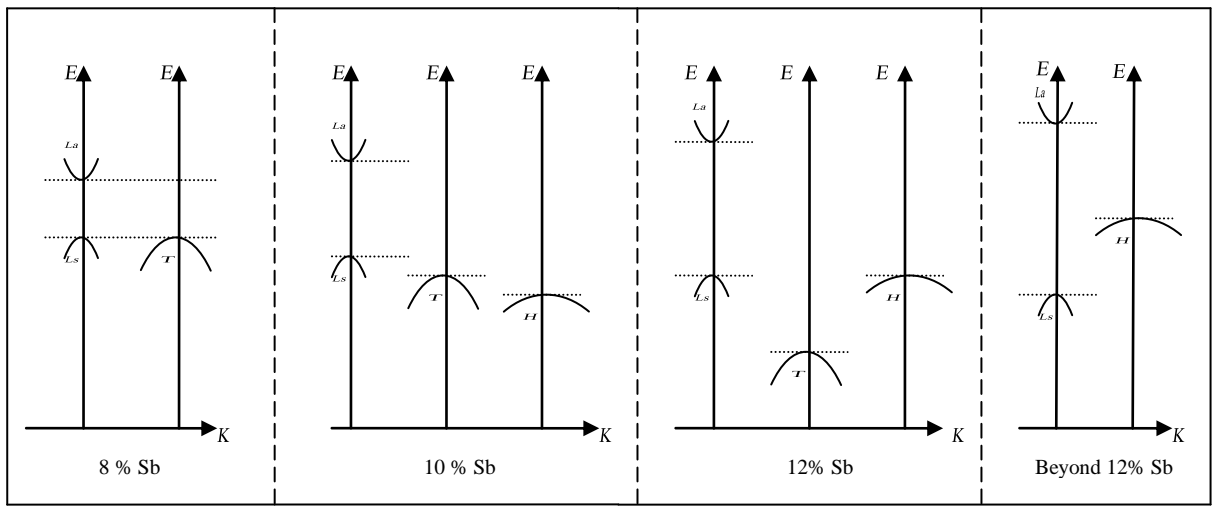